\documentclass[10pt,journal,letterpaper,compsoc]{IEEEtran}
\usepackage{times}
\usepackage[final]{graphicx}
\usepackage[reqno]{amsmath}
\usepackage{amsfonts}

\usepackage{times,amsmath,epsfig}
\usepackage{latexsym,amssymb}
\usepackage{cite}

\usepackage{graphicx}         
\usepackage{color}            
\usepackage{epic}             
\usepackage{eepic}            
\usepackage{rotating}         
\usepackage{type1cm}          
\usepackage{epsfig}

\newcommand{\beq}{\begin{equation}}
\newcommand{\enq}{\end{equation}}
\newcommand{\beqa}{\begin{eqnarray}}
\newcommand{\enqa}{\end{eqnarray}}
\newcommand{\beqn}{\begin{eqnarray*}}
\newcommand{\enqn}{\end{eqnarray*}}

\newcommand{\qed}{\hfill $\Box$}

\begin{document}

\title{Cognitive MAC Protocols for General Primary Network Models}

\author{Omar~Mehanna,~
        Ahmed~Sultan~
        and~Hesham~El Gamal
\IEEEcompsocitemizethanks{
\IEEEcompsocthanksitem O. Mehanna and A. Sultan are with the Wireless Intelligent Networks Center (WINC),
Nile University, Cairo, Egypt.\protect\\
E-mail: {omar.mehanna@nileu.edu.eg}, {asultan@nileuniversity.edu.eg}
\IEEEcompsocthanksitem H. El Gamal is with the Department
of Electrical and Computer Engineering, Ohio State University, Columbus, USA.\protect\\
E-mail: {helgamal@ece.osu.edu}
\IEEEcompsocthanksitem This work was presented in part at ICC'09.
\IEEEcompsocthanksitem This work was partially supported by the Egyptian NTRA and the National Science Foundation (NSF).
}
\thanks{}}


\markboth{Submitted to IEEE Transactions on Mobile Computing}%
{Shell \MakeLowercase{\textit{et al.}}: Bare Demo of IEEEtran.cls for Journals}

\IEEEcompsoctitleabstractindextext{%
\begin{abstract}
We consider the design of cognitive Medium Access Control (MAC) protocols enabling a secondary (unlicensed) transmitter-receiver pair to communicate over the idle periods of a set of primary (licensed) channels. More specifically, we propose cognitive MAC protocols optimized for both slotted and un-slotted primary networks. For the slotted structure, the objective is to maximize the secondary throughput while maintaining synchronization between the secondary pair and not causing interference to the primary network. Our investigations differentiate between two sensing scenarios. In the first, the secondary transmitter is capable of sensing all the primary channels, whereas it senses only a subset of the primary channels in the second scenario. In both cases, we propose blind MAC protocols that efficiently learn the statistics of the primary traffic on-line and asymptotically achieve the throughput obtained when prior knowledge of primary traffic statistics is available. For the un-slotted structure, the objective is to maximize the secondary throughput while satisfying an interference constraint on the primary network. Sensing-dependent periods are optimized for each primary channel yielding a MAC protocol which outperforms previously proposed techniques that rely on a single sensing period optimization.

\end{abstract}

}

\maketitle


\section{Introduction}
The radio spectrum resource is of fundamental importance
to wireless communication. Recent reports show that most
available spectrum has been allocated. However, most of licensed spectrum resources are under-utilized. This
observation has encouraged the emergence of dynamic and
opportunistic spectrum access concepts, where secondary (unlicensed)
users (SU) equipped with cognitive radios are allowed to
opportunistically access the spectrum as long as they do not
interfere with primary (licensed) users (PU). To achieve this goal, the
secondary users must monitor the primary traffic in order to
identify spectrum holes or opportunities which can be exploited to
transfer data \cite{Haykin}.

The main goal of a cognitive MAC protocol is
to sense the radio spectrum, detect the occupancy state of different
primary channels, and then opportunistically communicate
over unused channels (spectrum holes). Specifically, the cognitive MAC protocol should
continuously make efficient decisions on which channels to sense and
access in order to obtain the most benefit from the available
spectrum opportunities. Previous work on the design of cognitive MAC protocols has considered two distinct scenarios. In the first, the primary network is slotted (e.g., \cite{HangSu}, \cite{Capacity}, \cite{Zhao}, \cite{ZhaoWhitle}, \cite{Lifeng} and \cite{Motamedi}) whereas a continuous structure (un-slotted) of the primary channels is adopted in the second set of works (e.g., \cite{HyoilKim}, \cite{HyoilKim2}, \cite{OptimalSensing} and \cite{newZhao}). In this work we propose decentralized cognitive MAC protocols for each of the two models.

For the slotted structure, two cases are considered. The first
assumes that the secondary transmitter can sense all the available
primary channels before making the decision on which one to
access. The secondary receiver, however, does not participate in
the sensing process and {\em waits to decode} on only one
channel. This is the model adopted in~\cite{Capacity} and is intended to limit the
decoding complexity needed by the secondary receiver. In the
sequel, we propose an efficient algorithm that optimizes the
on-line learning capabilities of the secondary transmitter and
ensures perfect synchronization between the secondary pair. The
proposed protocol does not assume a separate control channel, and
hence, piggybacks the synchronization information on the same data
packet.

In the second scenario, the secondary transmitter can only sense a subset of the available
primary channels at the beginning of each time slot. This model
was studied in \cite{Zhao} where the optimal algorithm was obtained by formulating the problem as a Partially Observable Markov Decision Process (POMDP). Unfortunately,
finding a computationally efficient version of the optimal solution for this problem remains an elusive
task. By re-casting the problem as a restless multi-armed bandit problem, a near-optimal index policy was proposed in~\cite{ZhaoWhitle}. The authors of \cite{Zhao} and \cite{ZhaoWhitle}, however,
assumed that the primary traffic statistics (i.e., Markov chain
transition probabilities) were available {\em a-priori} to the
secondary users. Here, we develop {\bf blind} MAC protocols where
the protocol must learn the transition probabilities
on-line. This can be viewed as the Whittle index strategy
of~\cite{ZhaoWhitle} augmented with a similar learning phase to the
one proposed in~\cite{Lifeng} for the multi-armed bandit scenario.
Our numerical results show that the performance of this protocol
converges to that of the Whittle index strategy with known transition
probabilities~\cite{ZhaoWhitle}.

Under the un-slotted primary network set-up, we first assume that the SU radio can be tuned to any combination of the primary channels at the same time. This can be achieved by an Orthogonal Frequency Division Multiplexing (OFDM) technique with adaptive and selective allocation of OFDM subcarriers to utilize any subset of licensed channels at the same time. The SU aims at maximizing its throughput (\emph{i.e.,} maximizing the opportunities discovered and accessed in all primary channels) while imposing minimal interference to the primary network. A similar model was adopted in \cite{HyoilKim}, where the authors developed an optimal sensing period for each of the primary channels by optimizing the tradeoff between the sensing overhead resulting from frequent sensing of the channels and the missed opportunities in the primary channels due to infrequent sensing. However, it was assumed that if a primary transmission is resumed on a channel, the SU will discover this return, via the help of a Genie, and immediately evacuate the channel, thereby causing no interference to the primary transmissions. In this work, we relax this Genie-aided assumption and impose an interference/outage constraint on each primary channel. In \cite{OptimalSensing}, an optimal sensing period satisfying a primary network interference constraint was developed. However, the approximations made in~\cite{OptimalSensing} deviate considerably from the true values. More importantly, we show that by introducing two different sensing periods, a period if the channel is sensed free and a different period if the channel is sensed busy, the performance can be substantially improved. In particular, this performance improvement becomes more significant when there is a large difference between the expected time a primary channel is busy and the expected time it remains idle. Finally, we consider the scenario when the SU radio can be tuned to only one channel. A SU in this case shall try to access a primary channel as long as it is free. When this channel switches to busy, the SU shall search other primary channels until a free channel is identified. A similar model was adopted in \cite{HyoilKim2}, where an optimal sequence of primary channels to be sensed was proposed. This optimal sequence aimed at minimizing the average delay in finding a free channel. Here, we extend this work by finding the period a free channel shall be accessed in order satisfy an interference/outage constraint on the primary network, which was not considered in \cite{HyoilKim2}.

The rest of the paper is organized as follows. Section~\ref{model}
presents our modeling assumptions. The
proposed cognitive MAC protocols for slotted primary networks are developed in Section~\ref{slotted}, whereas
the un-slotted scenario is investigated in Section~\ref{unslotted}. Numerical results for our proposed strategies are reported in Section~\ref{numerical}. Finally, Section~\ref{conclusion} summarizes our conclusions.

\section{Network Model}\label{model}

\subsection{Primary Network}
We consider a primary network consisting of $N$ independent channels. The presence or absence of
primary users in each channel can be modeled as alternating time intervals of busy and free states
with random durations. For channel $i \in 1,2,\ldots , N$, we model the sojourn time of a busy period as a random variable $T_i^0$ with the probability density function (p.d.f.) $f_{T_i^0}(y), y > 0$. Similarly, the p.d.f. of the sojourn time in a free period is given as $f_{T_i^1}(x), x > 0$. Busy and free periods are assumed to be independent and identically distributed (i.i.d.). We also assume that busy and free periods are independent of each other. The state of channel $i$ at time $t$, $S_i(t)$, is equal to $1$ if the channel is free and $0$ if busy.

\subsection{Secondary Pair}
The SU can sense any of the primary channels in order to identify the presence of a PU. After sensing the channel, the SU applies the channel access strategies as described in Sections~\ref{slotted} and \ref{unslotted}. The performance of the sensing stage is limited by two types of errors. If the secondary transmitter decides that a free channel is busy, it will refrain from transmitting, and a spectrum
opportunity is overlooked. This is the false alarm situation,
which is characterized by the probability of false alarm $P_{FA}$. On
the other hand, if the detector fails to classify a busy channel as
busy, a miss detection occurs resulting in interference with
primary user. The probability of miss-detection is denoted by
$P_{MD}$. If energy detection is used as a sensing method \cite{SensingTime}, the minimum required sensing time $T_s$ that satisfies a certain desired $P_{FA}$ and $P_{MD}$ is given by:
\begin{equation}\label{T_s}
    T_s = \frac{2}{f_s} \left[Q^{-1}(P_{FA}) - Q^{-1}(1-P_{MD}) \sqrt{1 + 2 \sigma}\right]^2 \sigma^{-2}
\end{equation}
Where, $f_s$ is the sampling frequency, $Q(x)$ is the tail probability of a zero-mean unit-variance Gaussian random variable and $\sigma$ is the PU signal-to-noise ratio \cite{SensingTime}.


\section{Slotted Primary Network}\label{slotted}

\begin{figure}
  \includegraphics[width=0.45 \textwidth]{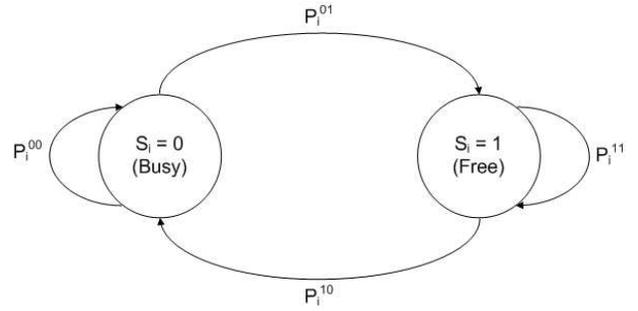}
  \caption{The Gilber-Elliot channel model.}\label{ChanState}
\end{figure}

In this section, we consider the case of discrete probability distributions for the free and busy periods.
We model the duration of these intervals (in terms of the number of time slots they occupy) as geometrically distributed
random variables. From the memoryless property of the geometric distribution, we can model the primary users' traffic in each channel by the two state Markov chain depicted in Figure~\ref{ChanState}. The channel state transition matrix of the Markov chain is given by
$P_i = \left[
   \begin{array}{cc}
     P^{00}_i & P^{01}_i \\
     P^{10}_i & P^{11}_i \\
   \end{array}
 \right]$.
We assume that $P_i$ remains fixed for a block of $T$ time slots.  We use $j$ to refer to the time-slot index
$j\in\{1,\cdots,T\}$.

We assume that the secondary transmitter can sense $L$
channels $(L \leq N)$ and can transmit on only one channel in each slot if the channel it
chooses to access is sensed to be free. In this section, $\bar{S}_i(j)$ denotes the
state of channel $i$ at time slot $j$ as sensed by the
transmitter, which might not be the actual channel state
$S_i(j)$. The secondary receiver does not participate in channel sensing and
is assumed to be capable of accessing only one
channel~\cite{Capacity}. This assumption is intended to limit the
decoding complexity needed by the secondary receiver. Another
motivation behind restricting channel sensing to the transmitter
is the potentially different sensing outcomes at the secondary
transmitter and receiver due to the spatial diversity of the
primary traffic which can lead to the breakdown of the secondary
transmitter-receiver synchronization. Overall, successful communication between the
secondary transmitter and receiver occurs only when: 1) they both
decide to access the same channel, and 2) the channel is sensed to
be free and is actually free from primary transmissions.

Our proposed cognitive MAC protocol can be decomposed into the following stages:
\begin{itemize}
    \item \textit{Decision stage}: The secondary transmitter decides which $L$ channels to sense. Also,
    both transmitter and receiver decide which channel to access.
    \item \textit{Sensing stage}: The transmitter senses the $L$ selected primary channels.
    \item \textit{Learning stage}: Based on the sensing results from the sensing stage, the transmitter updates the estimated primary channels' probability transition matrix $\hat{P_i}$.
    \item \textit{Access stage}: If the access channel is sensed to be free, a data packet is transmitted to the secondary
    receiver. This packet contains the information needed to sustain synchronization between secondary
    terminals, and hence, synchronization does not require a dedicated control channel. The length of the packet is assumed to be large enough such that the loss of throughput resulting from the synchronization overhead is marginal.
    \item \textit{ACK stage}: The receiver sends an ACK to the transmitter upon successful reception of sent data.
\end{itemize}

\subsection{Full Sensing Capability: $L = N$}

In this subsection we assume that the secondary transmitter can
sense all $N$ primary channels at the beginning of each time slot. Since the receiver doesn't participate in sensing, and in order to sustain the transmitter-receiver synchronization, the secondary pair must share the same variables which are used to decide upon the channel to be accessed. We refer to $ \Omega(j) = [\omega_1(j),\cdots,\omega_N(j)]$ as the belief vector at slot $j$, where $\omega_i(j)$ is the probability that $S_i(j) = 1$. Given the sensing outcomes in slot $j$, the belief state in slot $j+1$ can be obtained recursively as follows:
\begin{equation}
  \omega_i(j+1) =\begin{cases} P_{01}^i & \text{if}  \quad S_i(j)=0 \\
   P_{11}^i & \text{if}  \quad S_i(j)=1  \\
 \omega_i(j) P_{11}^i + (1- \omega_i(j)) P_{01}^i & \text{if $i$ not sensed at $j$} \end{cases}
\end{equation}

Due to the different sensing roles between the secondary transmitter and receiver, we introduce the vector $\bar{\Omega}(j)$ as the common or shared belief vector between the secondary transmitter and receiver. The initial packet sent to the receiver includes estimates for the
transition probabilities, and $\bar{\Omega}(1)$. Once the initial communication is
established, the secondary transmitter and receiver implement the
same spectrum access strategy described below for $j \geq 1$.

\begin{enumerate}
  \item \emph{Decision:} At the beginning of time slot $j$, and using belief vector $\bar{\Omega}(j)$, the secondary transmitter and receiver decide to access channel:
      \begin{equation}
        i^*(j) = \arg \max\limits_{i = 1,\cdots, N} \left[\bar{\omega}_i(j) B_i\right]
      \end{equation}
  \item  \emph{Sensing:} The secondary transmitter senses all channels and captures the sensing vector
        $\Phi(j) = [\bar{S}_1(j),\cdots,\bar{S}_N(j)]$, where $\bar{S}_i(j)=1$
        if the {\it i}th channel is sensed to be free, and $\bar{S}_i(j)=0$ if it is found busy. $\bar{S}_i(j)$ might be different than ${S}_i(j)$ due to sensing errors. Note that the decision stage precedes the sensing stage in order to maintain the synchronization between the secondary terminals.
  \item \emph{Learning:} Based on the sensing results, the transmitter updates the estimates $\hat{P}^{01}_i$ and $\hat{P}^{11}_i$ for all primary
  channels as explained below.
  \item  \emph{Access:} If $\bar{S}_{i^*}(j) = 1$, the transmitter sends its data packet to the receiver. The packet includes $\Phi(j)$, $\hat{P}^{01}_i$ and $\hat{P}^{11}_i$.
      In addition, if the transmission at slot $j-1$ has failed, the transmitter sends $\Omega(j)$, which is the belief vector computed at the transmitter based on its observations.
      If the receiver successfully receives the packet, it sends an ACK back to the
      transmitter. Parameter $K_{i^*}(j)$ is equal to unity if an
      ACK is received by the transmitter, and zero otherwise. If the
      channel is free, the forward transmission and the feedback
      channel are assumed to be error-free.
  \item Finally, the transmitter and receiver update the common belief vector
  $\bar{\Omega}(j+1)$ such that:
\end{enumerate}

If $K_{i^*}(j)= 1$:
\begin{equation}\label{omega_bar}
    \bar{\omega}_i(j+1) = \begin{cases}
   \bar{P}^{11}_i & \text{if $i = i^*(j)$}\\
   \bar{A}_i\bar{P}^{11}_i +  \left(1-\bar{A}_i\right)\bar{P}^{01}_i & \text{if $i \neq i^*(j), \bar{S}_i(j) = 1$}\\
   \bar{C}_i\bar{P}^{11}_i +  \left(1-\bar{C}_i\right)\bar{P}^{01}_i   & \text{if $i \neq i^*(j), \bar{S}_i(j) = 0$}
       \end{cases}
\end{equation}

If $K_{i^*}(j)= 0$:
\begin{equation}
    \bar{\omega}_i(j+1) = \begin{cases}
   \bar{D}_i\bar{P}^{11}_i +  \left(1-\bar{D}_i\right)\bar{P}^{01}_i & \text{if $i = i^*(j)$} \\
    \bar{\omega}_i(j)\bar{P}^{11}_i + (1-\bar{\omega}_i(j))\bar{P}^{01}_i  & \text{if $ i \neq i^*(j)$}
    \end{cases}
\end{equation}

\noindent where:
\footnotesize
\begin{equation}
    \bar{A}_i = Pr(S_i(j) = 1 | \bar{S}_i(j) = 1) = \frac{(1-P_{FA})\bar{\omega}_{i}(j)}{(1-P_{FA})\bar{\omega}_{i}(j) + P_{MD}(1-\bar{\omega}_{i}(j))}
\end{equation}

\begin{equation}
    \bar{C}_i = Pr(S_i(j) = 1 | \bar{S}_i(j) = 0) = \frac{P_{FA}\bar{\omega}_{i}(j)}{P_{FA}\bar{\omega}_{i}(j) + (1-P_{MD})(1-\bar{\omega}_{i}(j))}
\end{equation}

\begin{equation}
    \bar{D}_i = Pr(S_{i^*}(j) = 1 | K_{i^*}(j) = 0) = \frac{P_{FA}\bar{\omega}_{i}(j)}{P_{FA}\bar{\omega}_{i}(j) + (1 - \bar{\omega}_{i}(j))}
\end{equation}

\normalsize

$\bar{P}^{01}_i$ and $\bar{P}^{11}_i$ are the most recent shared
estimates of {\it i}th channel transition probabilities.
Obviously, in case of perfect sensing, $\bar{A}_i=1$,
$\bar{C}_i=0$ and $\bar{D}_i=0$.

In addition, the transmitter computes another belief vector,
$\Omega(j+1)$, based on its observations:

If $K_{i^*}(j)= 1$, $\omega_i(j+1) = \bar{\omega}_i(j+1)$\\

If $K_{i^*}(j)= 0$:
\begin{equation}\label{omega}
\omega_i(j+1) =
\begin{cases}
   A_i\hat{P}^{11}_i +  \left(1-A_i\right)\hat{P}^{01}_i & \text{if $ i \neq i^*(j), \bar{S}_i(j) = 1$}\\
   C_i\hat{P}^{11}_i +  \left(1-C_i\right)\hat{P}^{01}_i   & \text{if $i \neq i^*(j), \bar{S}_i(j) = 0$}\\
   D_i\hat{P}^{11}_i +  \left(1-D_i\right)\hat{P}^{01}_i & \text{if $i = i^*(j)$}
\end{cases}
\end{equation}

\noindent where $A_i$, $C_i$, and $D_i$ are the same as
$\bar{A}_i$, $\bar{C}_i$ and $\bar{D}_i$ with
$\bar{\omega}_i(j)$ replaced by $\omega_i(j)$. Note that
$\bar{\Omega}(1)=\Omega(1)$, and $\Omega(j+1)$ differs
from $\bar{\Omega}(j+1)$ only when $K_{i^*}(j)= 0$. If
transmission succeeds at the {\it j}th time slot after one or more
failures, the transmitter and receiver set
$\bar{\Omega}(j)=\Omega(j)$ before computing
$\bar{\Omega}(j+1)$.

It is noted that although $\Omega_i(j)$ is the updated belief vector which is available to the transmitter, the transmitter and receiver use the degraded $\bar{\Omega}(j)$ in the decision stage instead, in order to maintain the synchronization.
So, as an analytical benchmark, we have the following upper-bound on
the achievable throughput in this scenario. Assuming that the delayed
side information of all the
primary channels' states $S_i^{(j-1)}$ is known to the secondary
receiver as well as the transmitter, the expected throughput per slot is given by:
\begin{equation}\label{R}
    \small{R = \sum\limits_{{S_N}=0}^{1} \cdots \sum\limits_{{S_2}=0}^{1} \sum\limits_{{S_1}=0}^{1}\left[\left(\prod\limits_{i=1}^{N}P_{S_i}\right)\left(\max\limits_{i}\left[P_{S_i1}B_i\right]\right)\right]}
\end{equation}

\noindent where $P_{S_i1} $ denotes the state transition probability for
channel $i$ from state $S_i = (0 , 1)$ to the free state, and $P_{S_i}$
is the Markov steady state probability of channel $i$ being free or
busy. The first term in the summation corresponds to the probability
that the $N$ channels are in one of the $2^N$ states, and the second
term represents the highest expected throughput given the current
joint state for the $N$ channels. The loss in throughput resulting from the use of $\bar{\Omega}(j)$ instead of $\Omega_i(j)$ in the decision is illustrated in Figure~\ref{fig2}.

Since we assume that traffic statistics on primary channels
($P_i$) are unknown to the secondary user {\em a-priori}, the
secondary user needs to estimate these probabilities. When
continuous observations of each channel are available, each
channel can be modeled as a hidden Markov model (HMM). An optimal
learning algorithm for HMM is described in~\cite{HMM} using which
the transition probabilities, $P_{FA}$, and $P_{MD}$ can be
estimated. However, we here adopt a simple Bayesian learning method. Assume
that $P^{01}_i$ and $P^{11}_i$ are random variables with distributions
$f^{01}_t(x)$ and $f^{11}_i(y)$ defined on $[0, 1]$; respectively. After sensing all the primary channels at the beginning
of each time slot, and depending on the previous state of the channel, the posterior
distribution of $P^{01}_i$ can be updated according to Bayes' rule; i.e.,

\begin{eqnarray}\label{bayes}
  f^{01}_{j+1}(x) | \{01\}_j &=& \frac{x f^{01}_j(x)}{\int_0^1 x f^{01}_j(x)} \\
  f^{01}_{j+1}(x) | \{00\}_j &=& \frac{(1-x) f^{01}_j(x)}{\int_0^1 (1-x) f^{01}_j(x)},
\end{eqnarray}

\noindent where the event $\{01\}_j$ represents the state transition from busy at time $j-1$ to free at time $j$. Also, the event $\{00\}_j$ represents the state transition from busy at time $j-1$ to busy at time $j$. The posterior distribution of $P^{11}_i$ can be updated similarly. In addition, after sensing all the primary channels at the beginning
of each time slot, the secondary transmitter shall keep track of the
following metrics for each channel:

\begin{itemize}
\item Number of state transitions from busy to busy:
\begin{equation}
    N^{00}_i(j) = \sum\limits_{l=1}^{j-1} (1-\bar{S}_i(l))(1-\bar{S}_i(l+1))
\end{equation}
\item Number of state transitions from busy to free:
\begin{equation}
    N^{01}_i(j) = \sum\limits_{l=1}^{j-1} (1-\bar{S}_i{(l)})\bar{S}_i(l+1)
\end{equation}
\item Number of state transitions from free to busy:
\begin{equation}
    N^{10}_i(j) = \sum\limits_{l=1}^{j-1} \bar{S}_i(l)(1-\bar{S}_i(l+1))
\end{equation}
\item Number of state transitions from free to free:
\begin{equation}
    N^{11}_i(j) = \sum\limits_{l=1}^{j-1} \bar{S}_i(l)\bar{S}_i(l+1)
\end{equation}
\end{itemize}

Thus, if we assume that at $j=0$, $P^{01}_i$ is uniformly distributed in $[0,1]$ (i.e., $f^{01}_0(x) = 1$), or in other words no prior information about $P^{01}_i$ is available, then using equation (\ref{bayes}) it can be shown that $f^{01}_j(x)$ satisfies the following Beta distribution:
\begin{eqnarray}\label{bayes2}
    \nonumber \lefteqn{f^{01}_j(x | N^{00}_i(j)= N^{00}; N^{01}_i(j) = N^{01}) = }\\
    & & \frac{(N^{00}+ N^{01}+1) !}{N^{01}! +  N^{00}!} x^{N^{01}}(1-x)^{N^{00}}
\end{eqnarray}

Finally, the expected value of $f^{01}_j(x)$, obtained from equation (\ref{bayes2}), gives the following best estimate for $P^{01}_i$ at time $j$:
\begin{equation}
    \hat{P}^{01}_i(j) = \int_0^1 x f^{01}_j(x) dx = \frac{N^{01}_i(j) + 1}{N^{00}_i(j)+ N^{01}_i(j) + 2}
\end{equation}

Using the same approach, the best estimate for $P^{11}_i$ at time $j$ is given by:
\begin{equation}
    \hat{P}^{11}_i(j) = \frac{N^{11}_i(j) + 1}{N^{11}_i(j)+ N^{10}_i(j) + 2}
\end{equation}

This learning strategy can be easily applied to the situation where the primary traffic statistics ($P_i$) changes with time. One simple idea is to consider only a fixed number of previous sensing samples in estimating $P_i$ (i.e., using a sliding window of samples for the estimation)

In order to share the channel transition probabilities between the
secondary transmitter and receiver, as dictated by the proposed strategy, any updates in the values of $N^{00}_i(j)$, $N^{10}_i(j)$, $N^{11}_i(j)$
and $N^{01}_i(j)$ must be sent within the transmitted
packet. If $K_{i^*}(j)=1$, the transmitter and receiver update
$\hat{P}^{01}_i(j)$ and $\hat{P}^{11}_i(j)$. Otherwise, the
transmitter only updates $N^{00}_i(j)$ , $N^{10}_i(j)$, $N^{11}_i(j)$
and $N^{01}_i(j) $, but uses the old values, available at the last
successful transmission, in the decision phase.

In a nutshell, the proposed algorithm uses the full sensing
capability of the secondary transmitter to decouple the
exploration (i.e., learning) task from the exploitation task.
After an ACK is received, both nodes use the common
observation-based belief vector to make the optimal access
decision. On the other hand, in the absence of the ACK, both nodes
cannot use the optimal belief vector in order to maintain
synchronization. In this case, the proposed algorithm opts for a
greedy strategy in order to {\em minimize the time between the two
successive} ACKs.

A final remark is now in order. Assuming that $P^{11}_i =
P^{01}_i$, the probability of channel $i$ being free, $P_{S_i = 1}$,
becomes independent of the previous state, i.e., $P_{S_i = 1} =
P^{11}_i = P^{01}_i$. In this case, the optimal strategy, assuming
that the transition probabilities are known, is for the secondary
transmitter to access the channel $i^* = \arg \max\limits_{i =
1,\cdots, N} \left[P_{S_i = 1}B_i \right]$ and the expected
throughput becomes: $R = \max\limits_{i = 1,\cdots, N} \left[P_{S_i =
1}B_i\right]$~\cite{Lifeng}. Assuming, however, that the
transition probabilities are unknown but both nodes know that
$P^{11}_i = P^{01}_i$, one can estimate each channel free
probability as $\hat{P}_{S_i = 1} = N^1_i(j)/j$, where $N^1_i(j)$ is the number of times channel $i$ was sensed to be free until time slot $j$. In
Section~\ref{numerical}, we quantify the value of this side
information by comparing the performance of this strategy with our
proposed universal algorithm that does not make any prior assumption about
the transition probabilities.

\subsection{The Restless Bandit Scenario: $L < N$}\label{l11}

Here, we assume that the secondary transmitter can
sense only a subset $L < N$ of the primary channels at the beginning of each time slot.
Obviously, the problem here is not as simple as the $L = N$ case since the secondary transmitter has to decide on which $L$ channels to sense at each time slot in addition to the channel the transmitter and receiver decide to access. This opportunistic spectrum access network can be modeled as a partially observable Markov
decision process (POMDP) where the channel sensing and access of a MAC
protocol correspond to a policy for this POMDP \cite{Zhao}. The design objective is to determine, in each slot, which channel to sense so that the expected total reward: $E\left[\sum\limits_{j=1,i\in N}^{T} R_i(j)\right]$ obtained in $T$ slots is maximized, where $R_i(j) = S_i(j) B_i$. It has been shown that for any $j$, the belief vector $ \Omega(j) =
[\omega_1(j),\cdots,\omega_N(j)]$ is a sufficient statistic for the design of the optimal action
in slot $j$ \cite{Guha}. A policy  for a POMDP is thus given by
a sequence of functions, each mapping from the current belief
vector to the sensing and access action to be taken in slot $j$. Unfortunately, finding the optimal policy for a general POMDP is
computationally prohibitive. In \cite{Zhao}, the authors proposed a reduced complexity strategy based on
the greedy approach that maximizes the per-slot throughput based
on already known information (i.e., at time slot $j$, transmit on channel $i^*(j) = \arg \max\limits_{i = 1,\cdots, N} \left[\omega_i(j) B_i\right]$). In a more recent work \cite{ZhaoWhitle}, the problem was re-casted as a
restless bandit problem.

Restless Multi-armed Bandit Processes (RMBP) are generalizations of the classical Multi-armed Bandit Processes (MBP). In the MBP, a player,
with full knowledge of the current state of each arm, chooses one out of $N$ arms to activate at
each time and receives a reward determined by the state of the activated arm. Only the activated
arm changes its state while the states of passive arms are frozen.
The objective is to maximize the long-run reward over the infinite horizon by choosing which
arm to activate at each time. The solution to the multi-armed bandit problem should be able to
maintain a balance between the \textit{exploration and exploitation} in
order to maximize the total reward. Whittle~\cite{Whittle} introduced the RMBP which allow multiple arms to be activated simultaneously and passive arms to also change states. In each slot, the user chooses one of two possible actions to make a particular arm passive or active. Whittle's index measures how attractive it is to activate an arm based on the concept of
subsidy for passivity. In other words, Whittle's index for a channel is the minimum subsidy that is needed to move a state from the active set to the passive set. In \cite{ZhaoWhitle}, the Whittle index $W_i(j)$ was obtained in closed form and was used to construct a more efficient medium access policy than the greedy approach. The $W_i(j)$ given in \cite{ZhaoWhitle} can be viewed as a combination of the immediate reward represented by $\omega_i(j)$ and a learning reward obtained from observing the state of the channel. Based on this Whittle index formulation, the maximum reward obtained at each time slot is given by $\mathbf{R}(j) = \sum\limits_{i=1}^{N} R_i(j)$, where:
\begin{eqnarray}
  R_i(j) =\begin{cases} W_{i}(j) & \text{if} \; i(j) = \arg \max\limits_{i = 1,\cdots, N} \left[W_i(j) B_i \right]\\
   W_i(j) - \omega_i(j) & \text{if}  \quad  i \in U_{L-1}(j)  \\
 0 &   \text{otherwise}\end{cases}
\end{eqnarray}
and $U_{L-1}(j)$ represents the set of $L-1$ channels with the largest $L-1$ values of $(W_i(j) - \omega_i(j))$ not including channel $i(j) = \arg \max\limits_{i = 1,\cdots, N} \left[W_i(j) B_i \right]$. Knowing the states of the set of channels $U_{L-1}(j)$ gives the largest observation reward (\emph{i.e., exploration}) which enhances future access decisions.

Here, we relax the assumption of the {\em a-priori} available
transition probabilities at the secondary transmitter/receiver.
This adds another interesting dimension to the problem since the
blind cognitive MAC protocol must now learn this statistical
information on-line in order to make the appropriate access
decisions. Inspired by the previous results of Lai {\em et al.}\ in
the multi-armed bandit setup~\cite{Lifeng}, we propose the
following simple strategy. The $N$ primary channels are divided into $\lceil \frac{N}{L} \rceil$ channel groups. Then, at the beginning of the $T$ slots, each of the of the channel groups are continuously monitored for an
initial learning period ($LP$) to get an estimate for $P^{11}_i$
and $P^{01}_i$. In summary, the proposed strategy works as follows:

\begin{enumerate}
  \item \emph{Initial learning period:} Each group of channels are continuously sensed for $LP$ time slots. At the end of the learning period, the transition probabilities are estimated as $ \hat{P}^{01}_i(j) = \frac{N^{01}_i(j) + 1}{N^{01}_i(j)+ N^{00}_i(j) + 2}$, $\hat{P}^{11}_i(j) = \frac{N^{11}_i(j) + 1}{N^{11}_i(j)+ N^{10}_i(j) + 2}$
  \item \emph{Decision:} At the beginning of any time slot ($j > \lceil \frac{N}{L} \rceil \times LP$), the secondary transmitter and receiver decide to access channel $i^*(j) = \arg \max\limits_{i = 1,\cdots, N} \left[W_i(j) B_i \right]$.
  \item \emph{Sensing:} The secondary transmitter senses the $L-1$ channels in the set $U_{L-1}(j)$ in addition to channel $i^*(j)$. The sensing vector  $\Phi_{L}(j)$ for the selected $L$ channels is captured
  \item \emph{Learning:} if $i^*(j) = i^*(j-1)$, update
    $N^{11}_i$, $N^{10}_i$, $N^{01}_i$, $N^{00}_i$, $\hat{P}^{11}_i$, and $\hat{P}^{01}_i$.
  \item  \emph{Access:} If $\bar{S}_{i^*}(j) = 1$, the transmitter sends its data packet to the receiver. The packet includes $\Phi_{L}(j)$, $\hat{P}^{01}_i$ and $\hat{P}^{11}_i$.
      In addition, if the transmission at slot $j-1$ has failed, the transmitter sends $\Omega(j)$
  \item The transmitter and receiver calculate $\bar{\Omega}$, while the transmitter calculates $\Omega$ :

\end{enumerate}

If $K_{i^*}(j)= 1$:
\footnotesize
\begin{equation}\label{omega_bar2}
    \bar{\omega}_i(j+1) = \begin{cases}
   \bar{P}^{11}_i & \text{if $i = i^*(j)$}\\
   \bar{A}_i\bar{P}^{11}_i +  \left(1-\bar{A}_i\right)\bar{P}^{01}_i & \text{if $i \in U_{L-1}(j) , \bar{S}_i(j) = 1$}\\
   \bar{C}_i\bar{P}^{11}_i +  \left(1-\bar{C}_i\right)\bar{P}^{01}_i   & \text{if $i \in U_{L-1}(j), \bar{S}_i(j) = 0$}\\
   \bar{\omega}_i(j)\bar{P}^{11}_i + (1-\bar{\omega}_i(j))\bar{P}^{01}_i & \text{if $i \notin U_{L-1}(j)$}
    \end{cases}
\end{equation}

\normalsize
$\omega_i(j+1) = \bar{\omega}_i(j+1)$\\

If $K_{i^*}(j)= 0$:
\footnotesize
\begin{equation}\label{omega_bar22}
    \bar{\omega}_i(j+1) = \begin{cases}
   \bar{D}_i\bar{P}^{11}_i +  \left(1-\bar{D}_i\right)\bar{P}^{01}_i & \text{if $i = i^*(j)$} \\
    \bar{\omega}_i(j)\bar{P}^{11}_i + (1-\bar{\omega}_i(j))\bar{P}^{01}_i  & \text{if $i \neq i^*(j)$}
    \end{cases}
\end{equation}

\begin{equation}\label{omega22}
\omega_i(j+1) =
\begin{cases}
   A_i\hat{P}^{11}_i +  \left(1-A_i\right)\hat{P}^{01}_i & \text{if $i \in U_{L-1}(j), \bar{S}_i(j) = 1$}\\
   C_i\hat{P}^{11}_i +  \left(1-C_i\right)\hat{P}^{01}_i & \text{if $i \in U_{L-1}(j), \bar{S}_i(j) = 0$}\\
   {\omega}_i(j)\hat{P}^{11}_i + (1-{\omega}_i(j))\hat{P}^{01}_i & \text{if $i \notin U_{L-1}(j)$}\\
   D_i\hat{P}^{11}_i +  \left(1-D_i\right)\hat{P}^{01}_i & \text{if $i = i^*(j)$}
\end{cases}
\end{equation}

\normalsize

\noindent where $\bar{P}^{11}_i$ and $\bar{P}^{01}_i$ are the
latest successfully shared $\hat{P}^{11}_i$ and $\hat{P}^{01}_i$
between the secondary transmitter-receiver pair. Finally,
$\bar{\Omega}(j+1)$ is used to update Whittle's index
$W_i(j+1)$ of each channel as detailed in \cite{ZhaoWhitle}.

In the case of time-independent channel states, i.e., $P^{11}_i =
P^{01}_i$, the problem reduces to the multi-armed bandit
scenario considered in~\cite{Lifeng}. The difference, here, is the
lack of the dedicated control channel, between the cognitive
transmitter and receiver, as assumed in~\cite{Lifeng}. Assuming $L=1$, the
following strategy, which is applied as soon as the initial
synchronization is established, avoids this drawback by ensuring
synchronization using the ACK feedback over the same data channel.
\begin{enumerate}
  \item \emph{Decision:} At the beginning of any time slot $j$, the secondary transmitter and receiver decide
  to access the channel $i^*(j) = \arg \max\limits_{i = 1,\cdots, N} \left[\gamma_i(j)
  B_i\right]$, where $\gamma_i(j)=\frac{X_i(j)}{Y_i(j)} +
  \sqrt{\frac{2lnj}{Y_i(j)}}$, $X_i(j)$ is the number of
  time slots where successful communication occurs on channel $i$,
  and $Y_i(j)$ is the number of time slots where channel $i$ is
  chosen to sense and access \cite{Lifeng}.
  \item \emph{Sensing:} The secondary transmitter senses channel
  $i^*(j)$.
  \item  \emph{Access:} If $\bar{S}_{i^*}(j) = 1$, the transmitter sends its data packet to the receiver. If the receiver successfully receives a packet, it sends an ACK back to the transmitter.
  \item The transmitter and receiver update the following:\\
        $Y_{i}(j+1) = Y_{i}(j)+1 $, if $i(j)=i^*(j)$\\
        $X_{i}(j+1) = X_{i}(j)+1 $, if $K_{i^*}(j) = 1 , i(j)=i^*(j)$\\
        $\gamma_i(j+1) = \frac{X_i(j+1)}{Y_i(j+1)} + \sqrt{\frac{2lnj}{Y_i(j+1)}}$
\end{enumerate}


\section{Un-Slotted Primary Network}\label{unslotted}

In this section, we consider the case of continuous probability distributions for the free and busy periods.
Whenever the channel enters the busy or free state, the time until the next state transition is governed by the continuous p.d.f. $f_{T_i^0}(y)$ or $f_{T_i^1}(x)$; respectively.
Without loss of generality, we will use the following exponentially distributed busy/free periods for each channel as an illustrative example:
\begin{eqnarray}
  f_{T_i^1}(x) &=& \lambda_{T_i^1} e^{- \lambda_{T_i^1} x} \\
  f_{T_i^0}(y) &=& \lambda_{T_i^0} e^{- \lambda_{T_i^0} y}
\end{eqnarray}

\noindent The channel utilization $u_i$ in this case is given by:
\begin{eqnarray}
u_i &=& \frac{E[T_i^0]}{ E[T_i^0] + E[T_i^1]}  \\
&=& \frac{\lambda_{T_i^1}}{(\lambda_{T_i^1}+\lambda_{T_i^0})}
\end{eqnarray}

We also assume that the SU is equipped with a single antenna that can be used for either sensing or transmission.

\subsection{Multiple Channel Access}

Here, the SU can transmit on any combination of the $N$ primary channels simultaneously. However, the SU can sense only one channel at a time. Thus in order to sense any channel, the transmission taking place on any other channels is paused till the end of the sensing event. The goal is to find the optimal access strategy that maximizes the throughput for SU while satisfying the PU intereference/outage constraints for each channel.

Since the SU depends only on sensing a channel at specific times to identify the channel's state, it cannot track the exact state transition of each channel. Hence, the free portion of time between the actual state transition from busy to free until the SU discovers this transition cannot be utilized. In addition, some free periods may remain undiscovered at all if sensing is infrequent. These \emph{Unexplored Opportunities} are quantified by $T_i^U$, which is defined as the average fraction of time during which channel \emph{i}'s vacancy is not discovered by the SU \cite{HyoilKim}. On the other hand, the transition of primary activity from free to busy on a channel utilized by the SU causes interference to the primary and secondary receivers until the SU realizes this transition. This \textit{Interference Ratio} is quantified by $T_i^I$, which is defined as the average fraction of time at which channel $i$ is at the busy state but interrupted by SU transmission. Finally, we note that blindly increasing the sensing frequency to reduce interference and discover more opportunities is not desirable because the SU must suspend the use of the discovered channel(s) when it senses other channels. This is due to the assumption that data transmission and sensing cannot take place at the same time with one antenna. Thus the \textit{Sensing Overhead} $T_i^O$ is defined as the average fraction of time during which channel $i$ discovered opportunities are interrupted due to the need for sensing any of the $N$ channels \cite{HyoilKim}. This trade-off will be captured in the construction of our objective function which is used to find the optimal sensing frequencies/periods. The proposed algorithm relies on the novel idea of using two sensing periods for each channel: free sensing period $T_i^{F}$ if $S_i(t) = 1$, and busy sensing period $T_i^{B}$ if $S_i(t) = 0$. Therefore, our optimization task is to identify the optimal sensing periods $T_i^{F*}$, $T_i^{B*}$ for each channel, that maximize the total throughput for the SU on the $N$ channels while satisfying the PU interference constraint on each channel.

The channel as seen by the SU can be modeled by a two state (free/busy) Markov chain, where the transition probabilities from the free or busy state to the free state are: $P_i^{11}(t) = P(S_i(t_s + t) = 1 | S_i(t_s)=1)$ and $P^{01}(t) = P(S_i(t_s + t) = 1 | S_i(t_s)=0)$, $t_s$ is the most recent sensing time. For exponentially distributed busy/free periods, $P_i^{11}(t) $ and $P_i^{01}(t)$ are given by \cite{HyoilKim}:

\begin{eqnarray}
      P^{11}_i(t) &=& (1-u_i) + u_i e^{- ( \lambda_{T_i^1} + \lambda_{T_i^0}) t} \\
      P^{01}_i(t) &=& (1-u_i)-(1-u_i)e^{-(\lambda_{T_i^1}+\lambda_{T_i^0})t }
\end{eqnarray}

\noindent The ratio of the average number of times the channel is sensed free to the total number of times the channel is sensed can be obtained from the steady state probability that the Markov chain is in the free state:
\begin{equation}\label{PSS}
    P_i^{SS} = \frac{P_i^{01}(T_i^{B})}{1-P_i^{11}(T_i^{F})+ P_i^{01}(T_i^{B})}
\end{equation}

\noindent In case of perfect sensing (i.e., $P_{FA}=0$ and $P_{MD}=0$ ), $P_i^{SS}$ represents the probability that the SU senses channel $i$ with sensing-dependent periods $T_i^{F}$ and $T_i^{B}$, and finds it free. In the presence of sensing errors, the probability of finding channel $i$ free is:
\begin{equation*}
    (1-P_{FA}) \cdot P_i^{SS} + P_{MD} \cdot (1-P_i^{SS})
\end{equation*}

\noindent Note that the average time between sensing events on channel $i$ is given by:
\begin{eqnarray}\label{Avg}
    \nonumber \mu_i &=& P_i^{SS} \left[(1-P_{FA})T_i^{F} + P_{FA} T_i^{B}\right]\\
    & &  + \quad (1-P_i^{SS})\left[P_{MD}T_i^{F} + (1-P_{MD})T_i^{B} \right]
\end{eqnarray}

We define the \emph{Secondary Utilization} $T_i^{SU}(T_i^{F},T_i^{B})$ as the expected fraction of time during which channel $i$ is sensed or utilized by the SU,
\begin{equation}\label{PFree}
    T_i^{SU}(T_i^{F},T_i^{B}) = \left[(1-P_{FA})  P_i^{SS} + P_{MD}  (1-P_i^{SS})\right]\frac{T_i^{F}}{\mu_i}
\end{equation}

The total SU un-interrupted transmission time is equivalent to the expected throughput that can be achieved by the SU on all $N$ channels, and is given by:
\begin{eqnarray}\label{Rate}
    \nonumber R &=& \sum\limits_{i=1}^{N} \left[T_i^{SU}(T_i^{F},T_i^{B}) - T_i^I(T_i^{F},T_i^{B})  - T_i^O (T_i^{F},T_i^{B}) \right]\\
     &=& \sum\limits_{i=1}^{N} \left[(1-u_i) - T_i^U (T_i^{F},T_i^{B}) - T_i^O (T_i^{F},T_i^{B}) \right]
\end{eqnarray}

\begin{figure*}
  \includegraphics[totalheight=.4 \textheight, width=1 \textwidth]{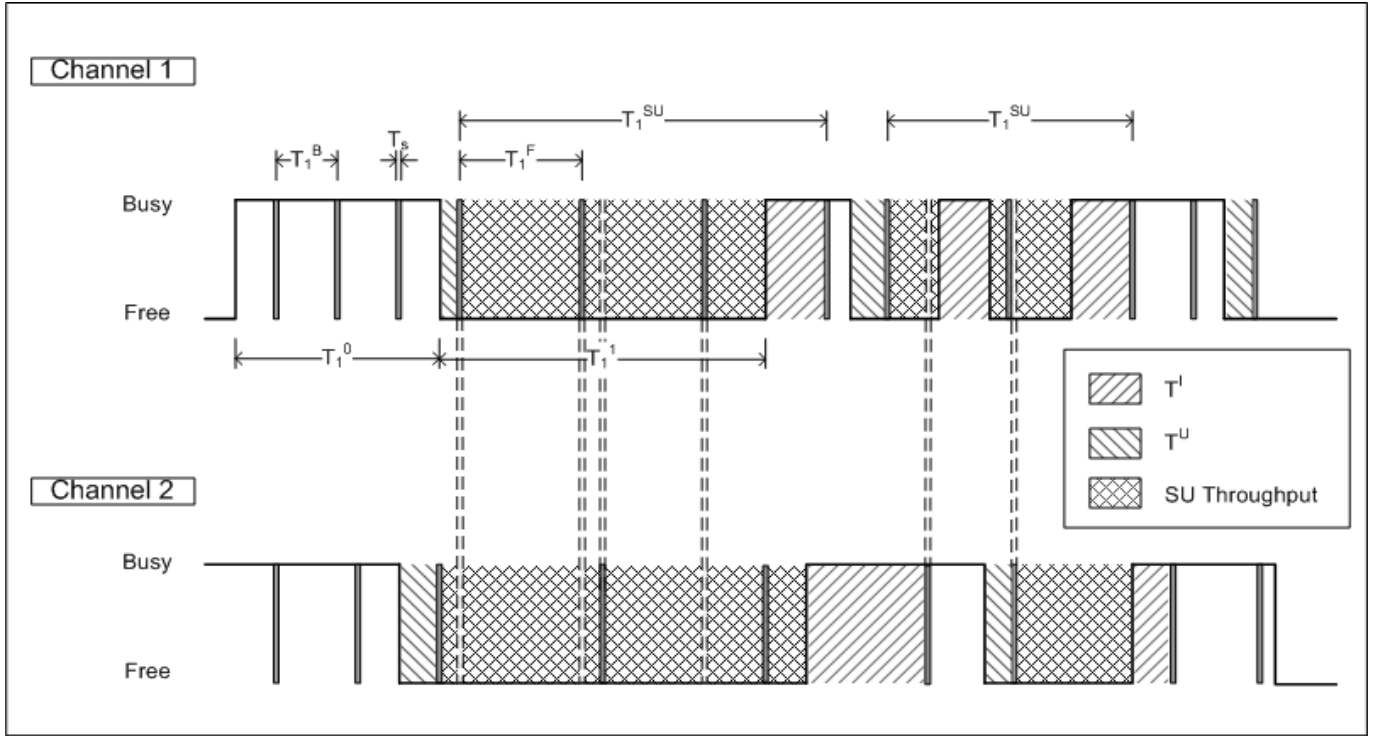}
  \caption{For a 2 channel primary network, this figure illustrates the periodic sensing using 2 periods per channel, the interference ratio, the unexplored opportunities, the sensing overhead, the secondary utilization, and the secondary achieved throughput on both channels.}
\label{Model} \end{figure*}

Figure \ref{Model} illustrates the sensing-dependent periods per channel, the interference ratio, the unexplored opportunities, the sensing overhead, the secondary utilization, and the secondary achieved throughput for a two primary channels model.

Now, in order to find expressions for the expected unexplored opportunities and interference ratio we need first to find  expressions for $\delta_i^1(t)$ and $\delta_i^0(t)$, which are defined as the expected time in which a channel is free during the time between $t_s$ and $t_s + t$ provided that $S_i(t_s) = 1$ or $S_i(t_s) = 0$; respectively. Based on the theory of alternating renewal processes, the remaining time $\tilde{x}$ for a channel to be in the same state from any sampling time $t_s$ can be shown to have the p.d.f. $\frac{1-F_{T_i^{1/0}}(x)}{E[T_i^{1/0}]}$ \cite{Cox}, where $F_{T_i^{1/0}}(x)$ is the c.d.f. of the free or busy period. Therefore, it can be easily shown that:
\begin{eqnarray}
  \nonumber \delta_i^1(t) &=& t \int_t^\infty \frac{1-F_{T_i^1}(x)}{E[T_i^1]} dx\\
  \label{delta1} & & +  \int_0^t \frac{1-F_{T_i^1}(x)}{E[T_i^1]} (x+\tilde{\delta}_i^0(t-x)) dx\\
  \label{delta0} \delta_i^0(t) &=&  \int_0^t \frac{1-F_{T_i^0}(y)}{E[T_i^0]} \tilde{\delta}_i^1(t-y) dy
\end{eqnarray}

\noindent where $\tilde{\delta}_i^1(t)$ and $\tilde{\delta}_i^0(t)$ are the same as $\delta_i^1(t)$ and $\delta_i^0(t)$ if the change in state happens exactly at $t_s$. That is,
\begin{eqnarray}
  \nonumber \tilde{\delta}_i^1(t) &=& t \int_t^\infty f_{T_i^1}(x) dx\\
  \label{delta_1} & & +  \int_0^t  f_{T_i^1}(x) (x+\tilde{\delta}_i^0(t-x)) dx\\
  \label{delta_0} \tilde{\delta}_i^0(t) &=& \int_0^t f_{T_i^0}(y) \tilde{\delta}_i^1(t-y) dy
\end{eqnarray}

Using Laplace transform, $\delta_i^1(t)$ and $\delta_i^0(t)$ for exponentially distributed busy/free periods can be obtained as: (see Appendix A for a complete derivation)
\begin{eqnarray}
   \delta^0_i(t)  & = & (1-u_i) \cdot \left(t+ \frac{ e^{-(\lambda_{T_i^0}+\lambda_{T_i^1})t} - 1}{ (\lambda_{T_i^0}+\lambda_{T_i^1}) }\right) \\
   \delta^1_i(t) & = &   t-u_i \cdot \left(t+ \frac{ e^{-(\lambda_{T_i^0}+\lambda_{T_i^1})t} - 1}{ (\lambda_{T_i^0}+ \lambda_{T_i^1}) }\right)
\end{eqnarray}

The unexplored opportunities $T^U_i$ can now be obtained as:
\begin{eqnarray}\label{UOPP1}
    \nonumber \lefteqn{T_i^U(T_i^{F},T_i^{B}) = }\\
    \nonumber & & (1-P_{MD})\cdot (1-P_i^{SS})\cdot \left(\frac{\delta_i^0(T_i^{B})}{\mu_i}\right) \\
    & &  + \quad P_{FA} \cdot P_i^{SS} \cdot \left(\frac{\delta_i^1(T_i^{B})}{\mu_i}\right)
\end{eqnarray}


\noindent Similarly, $T^I_i$ is given by:
\begin{eqnarray}\label{INTF1}
    \nonumber \lefteqn{T_i^I(T_i^{F},T_i^{B}) = }\\
    \nonumber & &(1-P_{FA}) \cdot P_i^{SS} \cdot \left(\frac{T_i^{F} - \delta_i^1(T_i^{F})}{\mu_i}\right) \\
    & &  + \quad P_{MD} \cdot (1- P_i^{SS}) \cdot \left(\frac{T_i^{F} - \delta_i^0(T_i^{F})}{\mu_i}\right)
\end{eqnarray}


Finally, the sensing overhead $T_i^O$ is given by:
\begin{equation}\label{SSOH}
	T_i^O (T_i^{F},T_i^{B}) = (T_i^{SU} - T_i^I) \sum_{j = 1}^N \left(\frac{T_s}{\mu_i}\right)
\end{equation}
\noindent It is worth noting that the first term $(T_i^{SU} - T_i^I)$ represents the average fraction of time channel $i$ is utilized by the SU without interference from the PU. This is the useful time that is interrupted by the need for sensing. The second term $\sum_{j = 1}^N \left(\frac{T_s}{\mu_i}\right)$ represents the aggregate sensing overhead given by the ratio of the sensing time to the average sensing sensing period.

In summary, given a maximum interference constraint per primary channel, $T_i^{Imax} $, our optimization problem can be expressed as follows:
\begin{eqnarray*}
	& & \text{Find: } T_i^{F*} , T_i^{B*}\\
    & & \text{that maximize: } \sum\limits_{i=1}^{N}  T_i^{SU}(T_i^{F},T_i^{B}) - T_i^I (T_i^{F},T_i^{B}) \\
    & & \qquad \qquad \qquad \qquad -  T_i^O (T_i^{F},T_i^{B})\\
    & & \text{subject to: } T_i^I(T_i^{F},T_i^{B}) \leq T_i^{Imax} \: ,\:  i = 1, \ldots , N
\end{eqnarray*}

\noindent This is equivalent to minimizing:
\begin{equation*}
    \sum\limits_{i=1}^{N} \left[ T_i^U (T_i^{F},T_i^{B}) + T_i^O (T_i^{F},T_i^{B}) \right]
\end{equation*}
subject to the same interference constraint. $T_i^{F*}$  and  $T_i^{B*}$ can be obtained numerically (as demonstrated by our numerical results in Section~\ref{numerical}).

\subsection{Single Channel Access}
Here, we assume that the SU can transmit on only one single channel. Our goals are 1) to find the optimal transmission period upon accessing any channel in order to satisfy an interference/outage constraint on the primary network and 2) to find the optimal sequence of primary channels to be sensed to minimize the average delay in finding a free channel.

When the SU finds a channel busy, it switches between channels and senses them until a free channel is found. Upon finding a vacant channel, the SU transmits for a period of $T_i^{F}$. In order to obtain the interference imposed on the primary transmission when the SU utilizes the channel for $T_i^{F}$, we note that this case is equivalent to setting $T_i^{B} = 0$ in the expressions derived in the previous section. Thus, the interference $T_i^I(T_i^{F})$ is given by:
\begin{eqnarray}\label{INTF2}
    \nonumber T_i^I(T_i^{F}) &=& (1-P_{FA})  \cdot \left(\frac{T_i^{F} - \delta_i^1(T_i^{F})}{T_i^{F}}\right) \\
    & &  + \quad P_{MD} \cdot \left(\frac{T_i^{F} - \delta_i^0(T_i^{F})}{T_i^{F}}\right)
\end{eqnarray}

\noindent Assuming error-free sensing, the interference $T_i^I(T_i^{F})$ for exponentially distributed busy/free periods becomes:
\begin{equation}\label{INTF_Mod}
    T_i^I(T_i^{F})=u_i \cdot \left(1+ \frac{ e^{-(\lambda_{T_i^0}+\lambda_{T_i^1})T_i^{F}} - 1}{ T_i^{F} \cdot (\lambda_{T_i^0}+\lambda_{T_i^1}) }\right)
\end{equation}

Let $T^{Imax}$ be the maximum fraction of outage/interference that the primary users can tolerate on all primary channels. Since only one channel is accessed at a given time, satisfying the interference constraint on each single channel is sufficient for ensuring that the total interference constraint is satisfied. Hence, assuming that the SU sensed channel $i$ to be free at time $t$ (i.e., $S_i(t)=1$), the sensing period $T_i^{F}$ for each channel must satisfy the constraint: $T_i^I(T_i^{F}) \leq T^{Imax}$. In order to maximize the SU throughput, the optimal sensing period for each channel $T_i^{F*}$ is $T_i^{F}$ which satisfies: $(T_i^I(T_i^{F}) = T^{Imax})$.

Now, we focus on the case when a channel is sensed to be busy. We need to find the optimal sequence of channels to sense in order to find a free channel as soon as possible, thereby minimizing the average delay in finding free channels. It is shown in \cite{HyoilKim2} that in order to minimize the average delay in finding a free channel, assuming all channels have the same capacity, the SU should attempt to access the channels in descending order of the channel index $\gamma_i$, where:

\begin{equation*}
    \gamma_i = \begin{cases} \frac{P^{11}_i(t-t_s)}{T_s} & \text{if}  \quad S_i(t_s) = 1\\
    \frac{P^{01}_i(t-t_s)}{T_s} &  \text{if} \quad  S_i(t_s) = 0 \end{cases}
\end{equation*}

In brief, the proposed strategy works as follows:
\begin{enumerate}
  \item Sense the $N$ channels in descending order of $\gamma_i$ until a free channel is found.
  \item Access the free channel $i$ for the calculated $T_i^{F*}$.
  \item When $T_i^{F*}$ ends, recalculate $\gamma_i$ for all channels, then repeat the previous steps.
\end{enumerate}

In order to synchronize the SU transmitter and receiver to the same channel while hopping, in the presence of sensing errors or in the case of spatially varying spectrum opportunities, we propose using \emph{Request To Send} (RTS) and \emph{Clear To Send} (CTS) packets. We assume that the packet duration of RTS and CTS is negligible compared to $T_i^{F}$ for all channels. If the transmitter senses a channel free, it will sends RTS. If the receiver gets the RTS, it replies with CTS, then transmission proceeds for $T_i^{F}$. If the receiver does not receive the RTS, it waits for the duration of CTS then switch to the next channel and when the transmitter does not receive CTS it also switches to the next channel. If the transmitter senses a channel busy, it waits for the duration of RTS, then switches to the next channel to sense. When the receiver does not receive the RTS, it shall switch to next channel. Thus synchronization between the secondary transmitter and receiver is always sustained. Note that in the case of multiple channel access, the secondary receiver is assumed to decode all primary channels, thus, no synchronization is needed.

%
%


\section{Numerical Results}\label{numerical}

\subsection{Slotted Primary Networks}

\begin{figure}
  \includegraphics[width=.5 \textwidth]{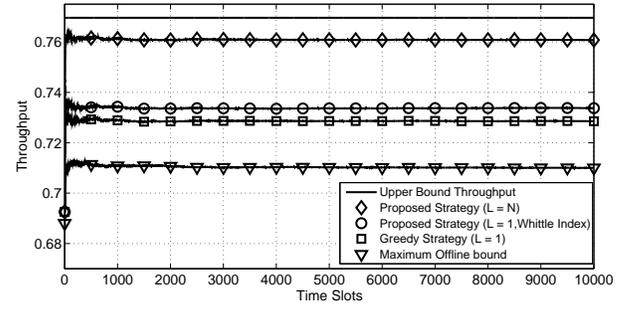}
  \caption{Throughput comparison between: the upper bound from equation (\ref{R}), the proposed blind strategy for $L=N$, the Whittle index strategy for $L=1$, the greedy strategy for $L=1$, and the maximum achievable offline bound.}
\label{fig2}\end{figure}

\begin{figure}
  \includegraphics[width=.5 \textwidth]{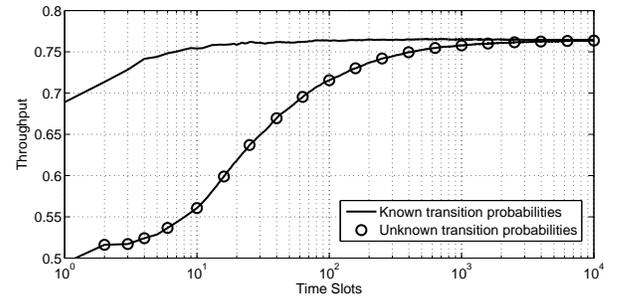}
  \caption{Throughput comparison between the proposed strategy for $(L = N)$ with and without known transition probabilities.}
\label{fig3} \end{figure}

\begin{figure}
  \includegraphics[width=.5 \textwidth]{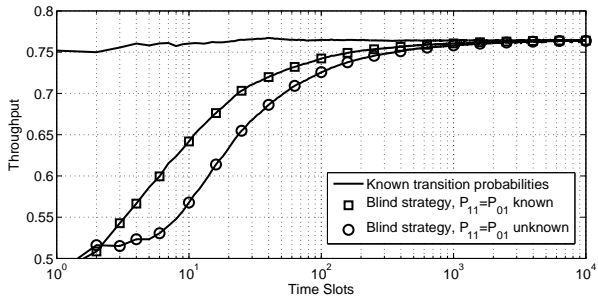}
  \caption{Throughput comparison for the blind cognitive MAC protocol (with and without the prior knowledge that $P^{11}_i = P^{01}_i$) and the genie-aided scenario.}
\label{fig4}\end{figure}

\begin{figure}
  \includegraphics[width=.5 \textwidth]{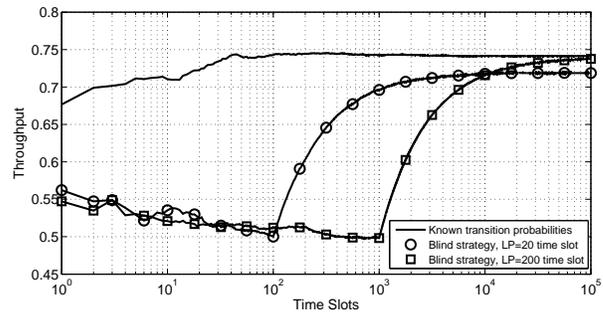}
  \caption{Throughput comparison between the proposed blind strategy for $(L = 1)$, when $LP = 20$ and $LP = 200$, and the genie-aided upper bound.}
\label{fig5}\end{figure}

In this subsection we present simulation results for the two
scenarios discussed earlier in Section~\ref{slotted}. Throughout this subsection, we assume
that the number of primary channels $N=5$, each with bandwidth
$B_i = 1$. The spectrum usage statistics of the primary network
were assumed to remain unchanged for a block of $T=10^4$ time
slots for Figures~\ref{fig2},~\ref{fig3}, and~\ref{fig4}, and for
a block of $T=10^5$ time slots for Figure~\ref{fig5}. The transition
probabilities for each channel, i.e., $P^{11}_i$ and $P^{01}_i$, were
generated uniformly between $0.1$ and $0.9$. The plotted results
are the average over $1000$ simulation runs. The discount factor
used to obtain Whittle's index is $0.9999$. In all our simulations, perfect sensing is assumed and the average
throughput per time slot is plotted.

Figure~\ref{fig2} reports the throughput comparison between the
different cognitive MAC strategies, all with prior knowledge about
the channels transition probabilities. First, it is noted that if the secondary transmitter can sense all $N$ channels neglecting the need for synchronization with the receiver, the probability of always finding a free channel $\approx 1$ for large $N$. However, since the strategies we consider sustain the secondary pair synchronization, a loss of throughput occurs as a price for synchronization. The topmost curve in Figure~\ref{fig2} gives the throughput obtained using equation (\ref{R}) assuming the delayed side information of all the primary channels' states is known to the secondary pair. The proposed strategy for the $L = N$ case is shown to achieve a throughput very close to the upper bound. Unsurprisingly, there is slight throughput gain offered by the full sensing capability as compared to the $L=1$ scenario. The difference, however, is not significant due to the constraint of maintaining synchronization which dictates choosing a channel to access followed by sensing. It is also seen from Figure~\ref{fig2} that the proposed strategies achieve a higher throughput than the protocol proposed in \cite{Capacity} where the primary transmitter and receiver are assumed to access the channels in a predetermined sequence, agreed upon a-priori, in which the channel with highest steady state probability of being free is always chosen for access.

Figure~\ref{fig3}
illustrates the convergence of the throughput of the proposed blind
strategy for $L=N$ to the informed case with
prior knowledge of the transition probabilities as $T$ grows. In
Figure~\ref{fig4}, we assume that $P^{11}_i = P^{01}_i$ for all
channels. It is shown that even if the secondary users are unaware
of this fact, and apply the proposed universal strategy, the achievable
throughput converges asymptotically to the achievable performance
when the fact that $P^{11}_i = P^{01}_i$ is known {\em a-priori},
albeit at the expense of a longer learning phase. Interestingly,
both strategies are shown to converge asymptotically to the genie-aided
upper bound (when the transition probabilities are known). It is noted in Figures~\ref{fig3} and \ref{fig4} that the proposed \emph{blind} strategies converge to the strategy with known transition probabilities in the range between $10^2$ and $10^3$ time slots.

Finally, Figure~\ref{fig5} demonstrates the tradeoff between the learning
time overhead in the \emph{blind} strategy of Section~\ref{l11} and the
final achievable throughput at the end of the $T$ slots assuming $L=1$. Clearly,
this figure supports the intuitive conclusion that for large $T$
blocks, one should use a longer learning phase in order to
maximize the steady state achievable throughput.

\subsection{Un-slotted primary channels structure}



\begin{figure}
  \includegraphics[width=.5 \textwidth]{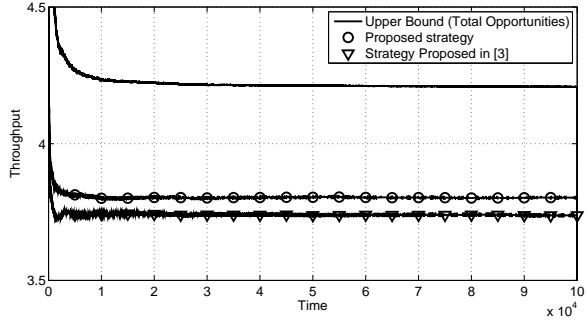}
  \caption{Performance comparison between the proposed multi-channel access strategy and the strategy proposed in \cite{HyoilKim} for an interference constraint $T_i^{Imax} = 0.25 \, u_i$}
\label{MultiChan1}\end{figure}

\begin{figure}
  \includegraphics[width=.5 \textwidth]{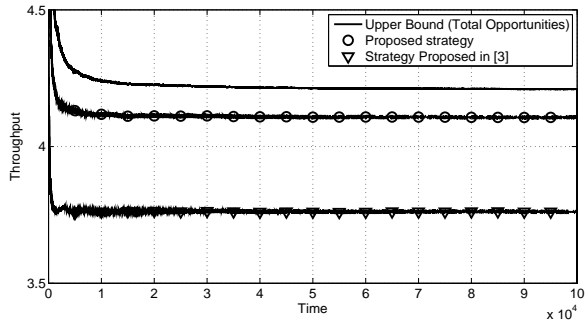}
  \caption{Performance comparison between the proposed multi-channel access strategy and the strategy proposed in \cite{HyoilKim}  for an interference constraint $T_i^{Imax} = 0.75 \, u_i$}
\label{MultiChan2}\end{figure}

In this subsection we present simulation results for the multiple channel access
scenario discussed in Section \ref{unslotted}. In Figures \ref{MultiChan1} and \ref{MultiChan2}, we compare the performance of the proposed strategy which adopts the sensing-dependent periods $T_i^{F}$ and $T_i^{B}$, and the strategy proposed in \cite{HyoilKim} which uses a single sensing period for each channel. In the simulations, we assume $N=5$ primary channels with exponentially distributed busy/free periods, where $\lambda_{T^1} = [0.2;0.17;0.15;0.13;0.11]$ and $\lambda_{T^0} = [1;0.9;0.8;0.7;0.6]$. Perfect sensing is assumed and the channel sensing duration is assumed to be $T_s=0.01$. The plotted results are the average over 100 simulation runs and the average throughput per time unit is plotted. The total available opportunities in the primary spectrum (upper bound on SU throughput) for the given values of $\lambda_{T^1}$ and $\lambda_{T^0}$ are: $\left(\sum\limits_{i=1}^5 (1-u_i)\right) = 4.205$ . Instead of the assumption that the SU will immediately detect returning PUs and evacuate the channel, which was used in \cite{HyoilKim}, we impose an interference constraint for each primary channel.

In Figure~\ref{MultiChan1}, the interference constraint for each channel $T_i^{Imax} = 0.25 \, u_i$. Under this assumption, our optimization method results in: $T^{F*} =[0.6133 ;   0.6800 ;   0.7637  ;  0.8714  ;  1.0148 ]$ and $T^{B*} = [ 0.3001  ;  0.3155  ;  0.3338  ;  0.3561  ;  0.3839]$. Using these values, the expected rate for the SU is given by $R=3.8068$. The optimization for the strategy proposed in \cite{HyoilKim} results in the single sensing period per channel $Tp_i^*$, where: $Tp^* = [ 0.6345  ;  0.7032  ;  0.7908  ;  0.9034  ;  1.0533]$  and an expected rate of $R=3.7531$. In Figure \ref{MultiChan2}, we set a more relaxed interference constraint for each channel $T_i^{Imax} = 0.75 \, u_i$. Our optimization method results in: $T^{F*} =[3.8847  ;  4.3127 ;   4.8462  ;  5.5318  ;  6.4457 ]$, $T^{B*} = [ 0.2793  ;  0.2950  ;  0.3135   ; 0.3359 ;   0.3637]$, and $R=4.1085$. The optimization for the strategy proposed in \cite{HyoilKim} results in: $Tp^* = [1.0444   ; 1.1035  ;  1.1403  ;  1.1886  ;  1.2532 ]$,  and $R=3.7731$. Overall, we can see from the two figures that the throughput of the proposed strategy outperforms that of the strategy proposed in \cite{HyoilKim} at different interference constraints. However, one can observe the following trend: As the interference constraint becomes more strict (i.e., $T_i^{Imax} \ll u_i$), more frequent sensing is required, resulting in a decreased throughput and a reduced advantage of the proposed strategy.


\section{Conclusions}\label{conclusion}

Assuming a slotted structure for the primary network, we proposed blind cognitive MAC protocols that do not require any prior knowledge about the statistics of the primary
traffic. Our work differentiated between two distinct scenarios, based on
the complexity of the cognitive transmitter. In the first, the
full sensing capability of the secondary transmitter was fully
utilized to learn the statistics of the primary traffic while
ensuring perfect synchronization between the secondary transmitter
and receiver in the absence of a dedicated control channel. The
second scenario focuses on a low-complexity cognitive transmitter
capable of sensing only a subset of the primary channels at the beginning of each time
slot. For this case, we proposed an augmented Whittle index MAC
protocol that allows for an initial learning phase to estimate the
transition probabilities of the primary traffic. Our numerical
results demonstrate the convergence of the blind protocols
performance to that of the genie-aided scenario where the primary
traffic statistic are known {\em a-priori} by the secondary
transmitter and receiver. For un-slotted primary networks, in order to maximize the secondary throughput achieved when being capable of transmitting on all primary channels simultaneously, we proposed a novel cognitive MAC protocol that utilize two sensing periods for each channel depending on the sensing outcome. Our numerical results show the superiority of the proposed protocol as compared to the protocols that rely on optimizing a single sensing period for each channel. On the other hand, when the SU can transmit on only one single channel, we show the optimal transmission period upon accessing any channel in order to satisfy an interference/outage constraint on the primary network.

\appendices
\section{Derivation of $\delta_i^1(t)$ and $\delta_i^0(t)$}
By taking the Laplace transforms for equations (\ref{delta1}), (\ref{delta0}), (\ref{delta_1}) and (\ref{delta_0}):

\begin{eqnarray}
\delta^0_i(s) &=& \frac{\mathbb{F}_{T_i^0}(s) \tilde{\delta}^1_i(s)}{ E[T_i^0]}\\
\delta^1_i(s) &=& \frac{1 - \mathbb{F}_{T_i^1}(s)}{s^2 . E[T_i^1]} + \frac{\mathbb{F}_{T_i^1}(s) \tilde{\delta}^0_i(s)}{E[T_i^1]}\\
\tilde{\delta}^0_i(s) &=& f_{T_i^0}(s) \tilde{\delta}^1_i(s)\\
\tilde{\delta}^1_i(s) &=& \frac{1 - f_{T_i^1}(s)}{s^2 } + \frac{f_{T_i^1}(s) \tilde{\delta}^0_i(s)}{E[T_i^1]}
\end{eqnarray}
\noindent where $\mathbb{F}_{T_i}(t) = 1-F_{T_i}(t)$.

Hence,
\begin{eqnarray}
 \delta^0_i(s) &=& \frac{\mathbb{F}_{T_i^0}(s)}{s^2 . E[T_i^0]}.\frac{1 - f_{T_i^1}(s) }{1- f_{T_i^1}(s) f_{T_i^0}(s)}\\
\delta^1_i(s) &=& \frac{1}{s^2 . E[T_i^1]}. \left[ 1 - \mathbb{F}_{T_i^1}(s). \frac{1- f_{T_i^0}(s) }{1- f_{T_i^1}(s) f_{T_i^0}(s)}\right]
\end{eqnarray}

And since for exponential distributions:
\begin{eqnarray*}
    f_{T_i}(s) &=& \frac{\lambda_{T_i}}{s+\lambda_{T_i}} \\
   \mathbb{F}_{T_i}(t) &=& e^{-\lambda_{T_i} t}\\
   \mathbb{F}_{T_i}(s) &=& \frac{1}{s+\lambda_{T_i}}
\end{eqnarray*}

We reach that:
\begin{eqnarray}
\nonumber \delta^0_i(s) &=& \frac{\frac{1}{s+\lambda_{T_i^0}}}{\frac{s^2}{\lambda_{T_i^0}}}\cdot \frac{1-\frac{\lambda_{T_i^1}}{s+\lambda_{T_i^1}}}{1-\frac{\lambda_{T_i^1}}{s+\lambda_{T_i^1}}\cdot \frac{\lambda_{T_i^0}}{s+\lambda_{T_i^0}}}\\
        &=& \frac{1}{s^2} \cdot \frac{\lambda_{T_i^0}}{s+ (\lambda_{T_i^0}+\lambda_{T_i^1})}\\
\nonumber \delta^0_i(t) &=& \int_{y=0}^t \left[\int_{x=0}^y \lambda_{T_i^0} \cdot e^{-(\lambda_{T_i^0}+\lambda_{T_i^1})x} dx\right] dy\\
\nonumber  &=& \frac{\lambda_{T_i^0}}{(\lambda_{T_i^0}+\lambda_{T_i^1})} \int_{y=0}^t [1-e^{-(\lambda_{T_i^0}+\lambda_{T_i^1})y}] dy\\
  &=& (1-u_i) \cdot \left(t+ \frac{ e^{-(\lambda_{T_i^0}+\lambda_{T_i^1})t} - 1}{ (\lambda_{T_i^0}+\lambda_{T_i^1}) }\right)
\end{eqnarray}

Similarly,
\begin{equation}
  \delta^1_i(t) =  t-u_i \cdot \left(t+ \frac{ e^{-(\lambda_{T_i^0}+\lambda_{T_i^1})t} - 1}{ (\lambda_{T_i^0}+\lambda_{T_i^1}) }\right)
\end{equation}


\begin{thebibliography}{10}

\bibitem{Haykin}
S.~Haykin, ``Cognitive radio: brain-empowered wireless communications," \emph{IEEE JSAC},
vol.~23, no.~2, pp.~201-220, February~2005.

\bibitem{HangSu}
H.~Su and X.~Zhang, ``Opportunistic MAC Protocols for Cognitive Radio," \emph{Proc. 41st Conference on Information
Sciences and Systems (CISS 2007)}, March~2007

\bibitem{HyoilKim}
H.~Kim and K.~Shin, ``Efficient Discovery of Spectrum Opportunities with MAC-Layer Sensing in Cognitive Radio Networks," \emph{IEEE Transactions on Mobile Computing},
vol.~7, no.~5, pp.~533-545, May~2008.

\bibitem{Capacity}
S.~Srinivasa, S.~Jafar and N.~Jindal, ``On the Capacity of the Cognitive Tracking Channel," \emph{IEEE International Symposium on Information Theory}, July~2006.

\bibitem{Zhao}
Q.~Zhao, L.~Tong, A.~Swami, and Y.~Chen, ``Decentralized Cognitive MAC for Opportunistic Spectrum Access in Ad Hoc Networks: A POMDP Framework," \emph{IEEE JSAC},
vol.~25, no.~3, pp.~589-600, April~2007.


\bibitem{HMM}
L.~Rabiner and H.~Juang, ``An introduction to hidden Markov models," \emph{IEEE ASSP Magazine},
vol.~3, no.~1, January~1986.

\bibitem{ZhaoWhitle}
K. Liu and Q. Zhao, ``Indexability of Restless Bandit Problems and Optimality of Whittle's Index for Dynamic Multichannel Access"
\emph{Submitted to IEEE Transactions on Information Theory},
November 2008.

\bibitem{Lifeng}
L.~Lai, H.~El-Gamal, H.~Jiang and H.~Poor, ``Cognitive Medium Access: Exploration, Exploitation and Competition," \emph{submitted to the IEEE Transactions on Networking}, October 2007.

\bibitem{Whittle}
P.~Whittle, ``Restless Bandits: Activity Allocation in a Changing World," \emph{Journal of Applied Probability}, vol.~25, 1988.

\bibitem{HyoilKim2}
H. Kim and K. Shin, ``Fast Discovery of Spectrum Opportunities in Cognitive Radio Networks",
\emph{Proceedings of the 3rd IEEE Symposia on New Frontiers in Dynamic Spectrum Access Networks (IEEE DySPAN)}
pp. 1-12, October 2008.

\bibitem{OptimalSensing}
Won-Yeol Lee and Ian. F. Akyildiz, ``Optimal Spectrum Sensing Framework for Cognitive Radio Networks,"
\emph{IEEE Transactions on Wireless Communications},
vol. 7, no. 10, October 2008.

\bibitem{SensingTime}
Zhi Quan, Shuguang Cui, H. Vincent Poor and H. Sayed, ``Collaborative wideband sensing for cognitive radios,"
\emph{IEEE, Signal Processing Magazine},
vol. 25, no. 6, pp. 60-73, November 2008.

\bibitem{Motamedi}
A. Motamedi and A. Bahai, ``Dynamic channel selection for spectrum sharing in unlicensed bands,"
\emph{European Transactions on Telecommunications and Related Technologies},
 2007.

\bibitem{newZhao}
Q. Zhao and J. Ye, ``When to Quit for a New Job: Quickest Detection of Spectrum Opportunities in Multiple Channels,"
\emph{in Proc. of IEEE Military Communication Conference (MILCOM)},
 November, 2008.

\bibitem{Cox}
D.R. Cox, ``Renewal Theory,"
 Butler and Tanner, 1967.

\bibitem{Guha}
S. Guha and K. Munagala, ``Approximation algorithms for partial-information based stochastic control with Markovian
rewards,"
\emph{in Proc. 48th IEEE Symposium on Foundations of Computer Science (FOCS)},
2007.

\end{thebibliography}
\end{document}